\documentclass[preprint,prd,nofootinbib,tightenlines,amsmath]{revtex4}
\usepackage{graphicx}
\usepackage{bm}

\begin{document}
\baselineskip=14pt \parskip=3pt

\vspace*{3em}

\title{Hidden Higgs Boson at the LHC and Light Dark Matter Searches
}

\author{Xiao-Gang He$^{1,2}$}
\author{Jusak Tandean$^{3}$}
\affiliation{$^1$INPAC, Department of Physics,
Shanghai Jiao Tong University, Shanghai, China \vspace*{1ex}
\\
$^2$Department of Physics and Center for Theoretical Sciences,
National Taiwan University, Taipei 106, Taiwan \vspace*{1ex}
\\
$^3$Department of Physics and Center for Mathematics and Theoretical Physics,
National Central University, Chungli 320, Taiwan
\vspace*{3ex} \\
}



\begin{abstract}
Recent LHC searches have not found a clear signal of the Higgs boson $h$ of the standard model
(SM) with three or four families in the mass range $m_h^{}=120$\,-\,600\,GeV.
If the Higgs had an unexpectedly large invisible branching ratio, the excluded $m_h^{}$ regions
would shrink.
This can be realized in the simplest weakly interacting massive particle dark matter~(DM) model,
which is the~SM plus~a~real gauge-singlet scalar field $D$ as the DM,
via the invisible mode~$h\to D D$.
Current~data~allow~this decay to occur for $D$-mass values near, but below, $m_h^{}/2$ and
those compatible with the light DM hypothesis.
For such $D$ masses, $h\to D D$ can dominate the Higgs width depending on~$m_h^{}$, and
thus sizable portions of the $m_h^{}$ exclusion zones in the SM with three or four families
may be recovered.
Increased luminosity at the LHC may even reveal a Higgs having SM-like visible decays still
hiding in the presently disallowed regions.
The model also accommodates well the new possible DM hints from CRESST-II
and will be further tested by improved data from future DM direct searches.
\end{abstract}

\maketitle

The hunt for the Higgs boson is an essential part of our effort to test the standard model~(SM).
Searches at the Tevatron~\cite{tevatron1} have ruled out the SM Higgs boson~$h$ with mass
$m_h^{}$ between 156 and 177~GeV at 95\% confidence level~(CL).
The latest results from the LHC have excluded most of the $m_h^{}$ range of
\,145\,-\,466\,GeV~\cite{atlas1,cms1} at~95\%~CL.
The LHC data are not yet sensitive to the region around \,$m_h^{}\sim120$\,GeV\, favored by
the SM fit to electroweak precision data~\cite{Baak:2011ze}.

The main production process for the SM Higgs at hadron colliders is the gluon fusion
\,$gg\to h$\, arising from a top-quark loop~\cite{Spira:1995rr}.
This mechanism is thus sensitive to new physics which can affect the loop-induced
\,$gg\to h$\, amplitude, such as extra heavy quarks.
Especially in the SM with four sequential fermion families~(SM4), the new heavy quarks can
enhance the Higgs production cross-section by up to {\footnotesize$\sim$}9
times~\cite{Li:2010fu} relative to that in the SM with three families~(SM3).
Reinterpreted in the SM4 context, the current LHC data then disallow $m_h^{}$ values
from 120 to~600\,GeV~\cite{atlas1,cms2}, which form a~sizable part of the range allowed by
precision data~\cite{Baak:2011ze}, while
Tevatron results exclude only~\,$m_h^{}=124$\,-\,286\,GeV~\cite{tevatron2}.

Future LHC searches may well discover a Higgs within the SM (either SM3 or SM4) preference.
However, if none is found between the LEP lower-bound and~1\,TeV,
physics beyond the SM may have to supply the explanation.
New physics could offset the heavy-quark contribution \mbox{to~\,$gg\to h$~\cite{Gunion:2011ww},}
reducing the $h$ production rate to below its SM value,
and/or the couplings involved in the decays on which Higgs searches rely could be less than
the corresponding couplings in the~SM,
as may occur in two-Higgs-doublet models~\cite{Gunion:2011ww}.
Alternatively, it is of course possible that an elementary Higgs does not exist at
all, if the electroweak symmetry breaking sector is a strongly correlated system~\cite{nohiggs},
or that the Higgs is simply too heavy for LHC data to reveal.
It may also be that the Higgs has an unexpectedly big branching-ratio~${\cal B}_{\rm inv}$
into invisible particles~\cite{He:2010nt,Keung:2011zc}.
Even if its production mechanisms and decays into visible channels are not altered,
a~larger-than-expected ${\cal B}_{\rm inv}$ will lower the event numbers of the Higgs decays
used in the LHC analyses, due to the increased rate of the invisible mode.
A~sufficiently high ${\cal B}_{\rm inv}$ will even make the present bounds on $m_h^{}$
disappear.
In this paper we focus on this last possibility and demonstrate that it may indeed be
realized in a~particle physics model of cold dark matter we call~SM+D.
This illustrates the potential deep connection between dark matter and Higgs physics.
The interplay between the two sectors might shine light on the still hidden elements in them.

The SM+D is one of the simplest models which can provide weakly interacting massive
particle~(WIMP) dark matter~(DM).
In addition to the SM particles, it has a real scalar field $D$ dubbed darkon which is a
singlet under the SM gauge group and acts as the~WIMP.
Beyond the SM (SM3 or SM4) part, the Lagrangian of the model has new renormalizable terms
given by~\cite{Silveira:1985rk,Burgess:2000yq}
\begin{eqnarray}  \label{DH}
{\cal L}_D^{} \,=\, \mbox{$\frac{1}{2}$}\partial^\mu D\,\partial_\mu^{}D
-\mbox{$\frac{1}{4}$}\lambda_D^{}D^4-\mbox{$\frac{1}{2}$}m_0^2D^2 - \lambda D^2 H^\dagger H ~, ~
\end{eqnarray}
where $\lambda_D^{}$,  $m_0^{}$, and $\lambda$  are free parameters and $H$ is the Higgs
doublet containing the physical Higgs field~$h$, in the notation of Ref.\,\cite{He:2010nt}
which gives some more details on the model.
Its DM sector has a small number of free parameters, only two of which, besides $m_h^{}$, are
relevant to our study: the Higgs-darkon coupling~$\lambda$ and the darkon mass
\,$m_D^{}=(m^2_0+\lambda v^2)^{1/2}$,\, where \,$v=246$\,GeV\, is
the Higgs vacuum expectation value.

\begin{figure}[b]
\, \includegraphics[width=94mm]{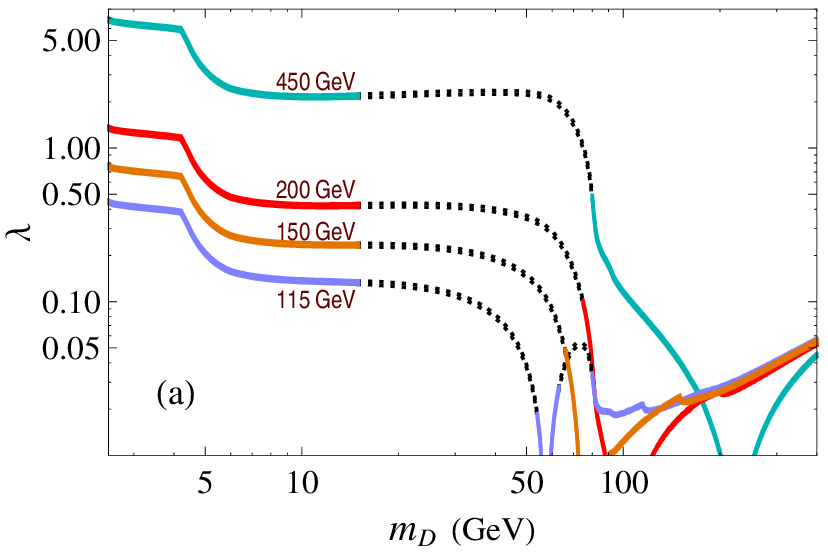}\vspace*{-1ex}\\
\includegraphics[width=97mm]{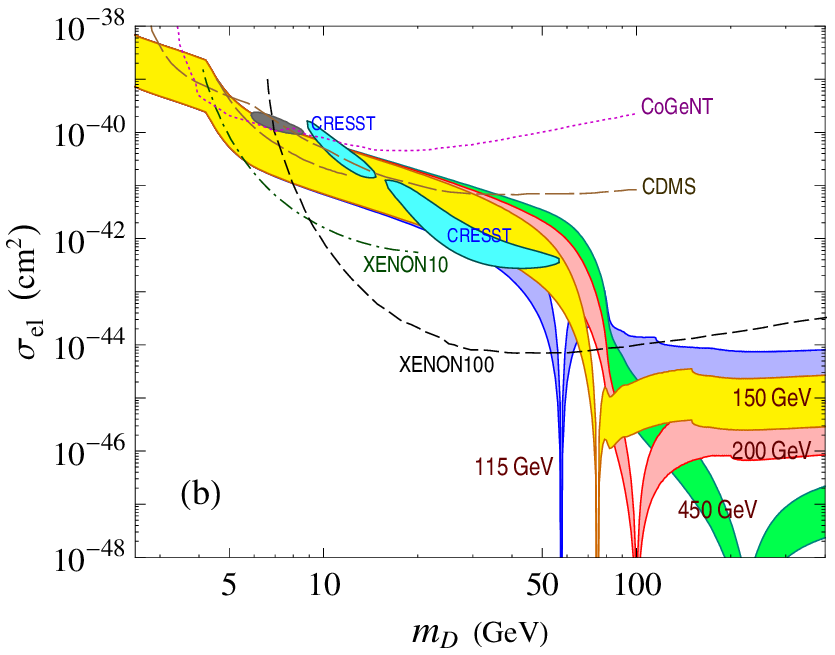} \, \,
\caption{(a)~Darkon-Higgs coupling $\lambda$ as a function of darkon mass $m_D^{}$ for
Higgs mass values $m_h^{}=115,150,200,450$~GeV  in~SM3+D.
(b)~The corresponding darkon-nucleon cross-section~$\sigma_{\rm el}^{}$, compared~to
{90\%-CL}~upper-limits from CoGeNT (magenta dotted curve)~\cite{Aalseth:2010vx},
CDMS (brown long-dashed curves)~\cite{Akerib:2010pv},
XENON10 (green dot-dashed curve)~\cite{Angle:2011th}, and
XENON100 (black short-dashed curve)~\cite{Aprile:2011hi}, as well as two (cyan)
areas representing the new CRESST-II result~\cite{cresst} and a dark-gray
patch fitting both DAMA/LIBRA~\cite{Bernabei:2010mq} and
CoGeNT~\cite{Aalseth:2010vx} signal data~\cite{Hooper:2010uy}.
The black-dotted sections of the curves~in~(a), and also in the following figures,
are disallowed by the direct-search limits in~(b) as discussed in
the text.\label{lambda-md}}  
\end{figure}

The darkon model can yield the required WIMP relic density by means of Higgs-mediated darkon
annihilation into kinematically allowed SM particles~\cite{Silveira:1985rk,Burgess:2000yq}.
Upon specifying $m_D^{}$ and~$m_h^{}$, one can extract $\lambda$ from the relic-density
number~\,$\Omega_D^{}h^2=0.1123\pm 0.0035$~\cite{wmap7}.
Applying the procedure given in~Ref.\,\cite{He:2010nt} to the~SM3+D case for
\,$2.5{\rm\,GeV}\le m_D^{}\le400$\,GeV\, and~some illustrative values of~$m_h^{}$, we present
the results in~Fig.\,\ref{lambda-md}(a), where the band widths reflect the relic-density range.
In the~SM4+D, the $\lambda$ results are mostly somewhat lower than their SM3+D counterparts,
by no more than {\footnotesize$\sim$}20\%, similarly to what was found in~Ref.\,\cite{He:2010nt}.
The reason for the decrease is that the Higgs total width in the~SM4 is enlarged relative to
that in the~SM3, mainly due to the rate of \,$h\to gg$\, being enhanced by the new heavy
quarks~\cite{He:2010nt}.

A number of underground experiments have been performed to detect WIMP DM directly
by looking for the recoil energy of nuclei due to the scattering of a~WIMP off
\mbox{a~nucleon~\cite{Bernabei:2010mq,Aalseth:2010vx,Akerib:2010pv,Angle:2011th,Aprile:2011hi,
cresst,Gelmini:2011xz}.}
The acquired data impose additional constraints on the parameter space of the darkon \mbox{models}.
The relevant observable is the spin-independent cross-section $\sigma_{\rm el}^{}$
of the darkon-nucleon elastic collision via $t$-channel
Higgs-exchange~\cite{Silveira:1985rk,Burgess:2000yq}.
Thus to compute $\sigma_{\rm el}^{}$ requires knowing not only~$\lambda$, but also
the Higgs-nucleon coupling~$g_{NNh}^{}$.
We again follow~Ref.\,\cite{He:2010nt}, but here employ a~range of $g_{NNh}^{}$ to account for
its substantial uncertainty arising from its dependence on the pion-nucleon sigma term
$\sigma_{\pi N}^{}$ which is not well determined~\cite{Ellis:2008hf}.
Since phenomenological analyses yield
\,\mbox{$36{\rm\,MeV}\le\sigma_{\pi N}^{}\le71$\,MeV~\cite{Gasser:1990ce},} while lattice calculation
results have a wider spread from \,{\footnotesize$\sim$}15 to~90~MeV~\cite{Young:2009ps},
we can reasonably take \,$30{\rm\,MeV}\le\sigma_{\pi N}^{}\le80$\,MeV.\,
With the aid of formulas from Refs.\,\cite{He:2010nt,sm+d},
this translates into \,$0.0011\le g_{NNh}^{\rm SM3}\le0.0032$\, and
\,$0.0016\le g_{NNh}^{\rm SM4}\le0.0033$.

We show in Fig.\,\ref{lambda-md}(b) the calculated $\sigma_{\rm el}^{}$ in the SM3+D for
the same choices of $m_D^{}$ and $m_h^{}$ as in~Fig.\,\ref{lambda-md}(a).
Also shown are results of the latest direct-searches for~DM,
including CRESST-II which has reported fresh possible WIMP hints~\cite{cresst}.
Evidently the $g_{NNh}^{}$ uncertainties can make $\sigma_{\rm el}^{}$ vary by up to
an order of magnitude~\cite{Ellis:2008hf}.  Nevertheless, including them leads to a~more
complete picture of how the data confront the darkon model.
In the SM4+D case, which is not shown, the majority of the predictions for~$\sigma_{\rm el}^{}$
are higher by~{\footnotesize$\sim$}50\%, and varying somewhat less, than their SM3+D counterparts.

Comparing the SM3+D predictions for $\sigma_{\rm el}^{}$ to the experimental bounds
in~Fig.\,\ref{lambda-md}(b), one can see that darkon masses
$\mbox{\small$\gtrsim$}\,15$\,GeV\, up to {\footnotesize$\sim$}80\,GeV\, are disallowed
except around~\,\mbox{$m_D^{}\sim m_h^{}/2$.}  We remark that these dips near
\,$m_D^{}=m_h^{}/2$\, are a common feature of $\sigma_{\rm el}^{}$ curves~\cite{sm+d}.
Specifically, we find
\,$54{\rm\,GeV}\,\mbox{\small$\lesssim$}\,m_D^{}\,\mbox{\small$\lesssim$}\,63$\,GeV\, is
allowed for \,$m_h^{}=115$\,GeV,\, and also
\,$m_D^{}\,\mbox{\small$\gtrsim$}\,66$, $75,80\rm~GeV$\,~for
\,$m_h^{}=150,200,450$~GeV,\, respectively. Their counterparts in the  SM4+D  are
\,$54{\rm\,GeV}\,\mbox{\small$\lesssim$}\,m_D^{}\mbox{\small$\,\lesssim\,$}62$\,GeV\,
for \,$m_h^{}=115$\,GeV\, and \,$m_D^{}\,\mbox{\small$\gtrsim$}\,67,78,81$~GeV\, for
\,$m_h^{}=150,200,450$~GeV,\, respectively.
More generally, the direct searches to date have not yet probed the darkon model much
beyond~\,$m_D^{}\sim80$\,GeV.\,

For lighter darkons, it may seem from Fig.\,\ref{lambda-md}(b) that almost all masses
down to \,$m_D^{}\sim5$\,GeV\, are already excluded by the CDMS,
XENON10, and XENON100 limits.
However, their results for WIMP masses $\mbox{\small$\lesssim$}\,$15\,GeV have been
seriously disputed in the literature~\cite{Gelmini:2011xz,Hooper:2010uy,Collar:2011kf}.
Furthermore, other recent searches~by DAMA/LIBRA~\cite{Bernabei:2010mq},
CoGeNT~\cite{Aalseth:2010vx}, and CRESST-II~\cite{cresst} have turned up
potential evidence for DM under 30\,GeV.
In particular, the excess events newly observed at CRESST-II may have been
caused by WIMPs of mass about 12 or 25~GeV~\cite{cresst}.
Pending a general consensus on this matter, it is not impossible that the DM masses
suggested by DAMA/LIBRA, CoGeNT, or CRESST-II are still viable.
One should therefore keep an open mind that this light-WIMP region is not totally ruled out.
It is then interesting to see that the SM3+D predictions in Fig.\,\ref{lambda-md}(b) overlap
well with the 2$\sigma$-confidence (cyan) areas compatible with the CRESST-II
result~\cite{cresst}, especially the lower mass region around~12\,GeV.
The predictions also cover part of the dark-gray area that can fit both the DAMA/LIBRA and
CoGeNT data at 99\%\,CL according to Ref.\,\cite{Hooper:2010uy}.
The same can be said of the corresponding results in the~SM4+D.

For even lower masses, the available data on $B$-meson decay
$B\to K\mbox{$\not{\!\!E}$}$ and kaon decay $K\to\pi\mbox{$\not{\!\!E}$}$ with
missing energy $\mbox{$\not{\!\!E}$}$ imply stringent restrictions excluding most of the
\,$m_D^{}<\bigl(m_B^{}-m_K^{}\bigr)/2\simeq2.4$\,GeV\, region~\cite{Bird:2004ts,He:2010nt}.
In contrast, bounds on \,$B\to\mbox{$\not{\!\!E}$}$\, and the bottomonium
decay \,$\Upsilon\to\gamma\mbox{$\not{\!\!E}$}$\, are still too weak~\cite{Yeghiyan:2009xc}
to probe higher masses up to~\,$m_D^{}\sim m_\Upsilon^{}/2\sim5$\,GeV.\,

Based on the preceding considerations, we regard~the range
\,$2.5{\rm\,GeV}\le m_D^{}\le15$\,GeV\, as still viable~in~the SM3+D and~SM4+D.
It accommodates the WIMP masses hinted at by CoGeNT as well as the smaller values
of those suggested by DAMA/LIBRA and CRESST-II.
The various allowed ranges of $m_D^{}$ discussed above are depicted for the SM3+D
in~Fig.\,\ref{lambda-md}(a), where the black-dotted sections of the $\lambda$ curves
are disallowed.

\begin{figure}[b] 
\includegraphics[width=77mm]{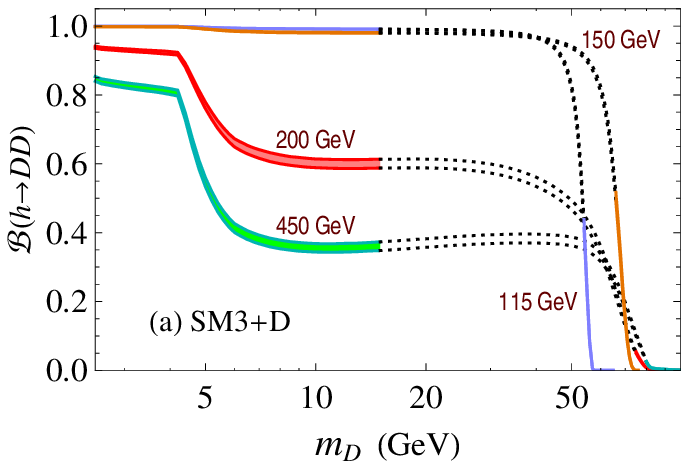} \, \,
\includegraphics[width=77mm]{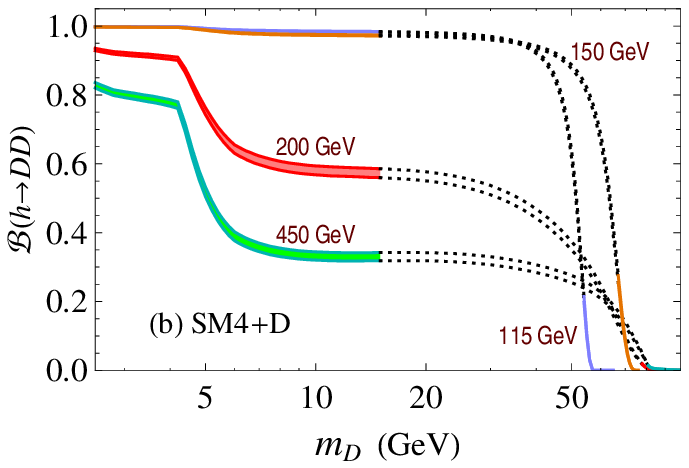}
\caption{Branching ratio of \,$h\to DD$\, as a function of $m_D^{}$ in (a)~SM3+D and (b)~SM4+D
for \,$m_h^{}=115,150,200,450$~GeV.\label{br-md}}
\end{figure}

Now if \,$m_h^{}>2m_D^{}$,\, the branching ratio of the invisible decay \,$h\to DD$\,
is \,${\cal B}(h\to DD)=\Gamma(h\to DD)/\Gamma_h^{\rm SM+D}$,\, where
\,$\Gamma(h\to DD)=\lambda^2v^2\bigl(1-4m_D^2/m_h^2\bigr){}^{1/2}/(8\pi m_h^{})$\, and
\,$\Gamma_h^{\rm SM+D}=\Gamma_h^{\rm SM}+\Gamma(h\to DD)$\, includes the Higgs total
width $\Gamma_h^{\rm SM}$ in the SM3 or SM4 without the darkon.
To illustrate how large ${\cal B}(h\to DD)$ can~be, we use the $\lambda$ values obtained
earlier to draw the plots in~Fig.\,\ref{br-md}, where the black-dotted areas are again excluded.
Obviously, in the viable $m_D^{}$ zones of the SM3+D or SM4+D the additional process
$h\to DD$ can greatly enhance
the Higgs invisible branching ratio \,${\cal B}_{\rm inv}^{}\simeq{\cal B}(h\to DD)$.\,
Needless to say, this implies potentially significant changes to the Higgs branching ratios
assumed in LHC analyses~\cite{others}.

The impact of the enlarged ${\cal B}_{\rm inv}^{}$ on Higgs searches can be quantified
in a different way.
Since the darkon has no gauge interactions or mixing with the Higgs,
the rates of Higgs decays into $\gamma\gamma,\tau^+\tau^-,b\bar b,WW^{(*)}$,
and $ZZ^{(*)}$, which are employed in LHC searches~\cite{atlas1,cms1,cms2},
are not modified in the SM+D with respect to the~SM~alone.
It follows that their branching ratios in the SM+D are all subject to the same reduction
factor~\cite{Burgess:2000yq}
\begin{eqnarray}
{\cal R} ~=~ \frac{{\cal B}\bigl(h\to X\bar X\bigr)}{{\cal B}\bigl(h\to X\bar X\bigr)_{\rm SM}}
~=~ \frac{\Gamma_h^{\rm SM}}{\Gamma_h^{\rm SM}+\Gamma(h\to DD)} ~.
\end{eqnarray}
Since the $gg\to h$ expectation is unchanged by the darkon's presence,
the cross-section of \,$gg\to h\to X\bar X$\, is then decreased by the same factor $\cal R$,
as are the cross sections of other Higgs production modes.
Hence the assumed event rate for each production channel in the SM Higgs searches would be
overestimated by \,$1/\cal R$\, times.

\begin{figure}[t] 
\includegraphics[width=90mm]{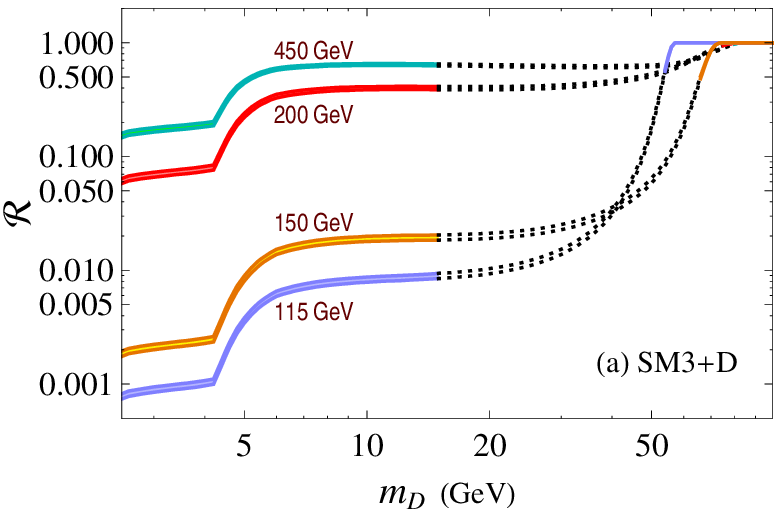} \vspace*{-1ex} \\
\includegraphics[width=90mm]{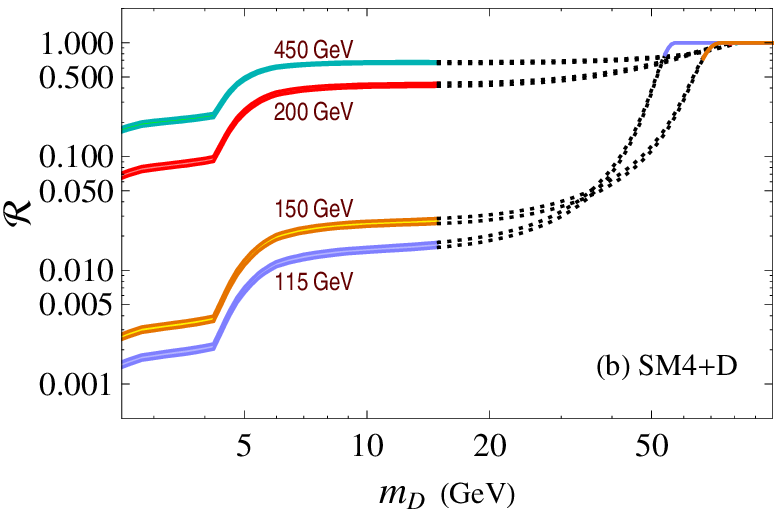} \vspace*{-1ex}
\caption{Reduction factor $\cal R$ as a function $m_D^{}$  in (a)~SM3+D and (b)~SM4+D for
\,$m_h^{}=115,150$, $200,450$~GeV.\label{rfactor}}
\end{figure}

In Fig.\,\ref{rfactor} we plot $\cal R$ for the same $m_D^{}$ and $m_h^{}$ choices as
in~Fig.\,\ref{br-md}.  These graphs indicate that the darkon effect can suppress the Higgs
branching ratios into SM particles by up to 3 orders of magnitude.  More precisely, the values
of ${\cal R}$ in the viable regions of~$m_D^{}$ are collected in~Table~\ref{rvalues}.
One notices that the two allowed regions of $m_D^{}$ lead to two distinct ranges of $\cal R$
in each $m_h^{}$ case, the gap between them narrowing as $m_h^{}$ increases.
Moreover, $\cal R$~at a particular $m_D^{}$ rises drastically right after $m_h^{}$ exceeds
$2m_W^{}$ and the channel \,$h\to WW$\, is fully open, which quickly builds
up~$\Gamma_h^{\rm SM}$.
It is worth remarking, in addition, that the upper limits of $\cal R$ in the viable
low-$m_D^{}$ region would not change much if its maximum value were increased from
15\,GeV to 20\,GeV (or even 30\,GeV).

With our $\cal R$ examples, we can explore how the darkon effect may alter the LHC
limits on~$m_h^{}$.
Since the determination of the $m_h^{}$ exclusion zones is based on the measured upper-limit
on the Higgs production cross-section at the $pp$ collider divided by its SM expectation,
$\sigma/\sigma_{\rm SM}^{}$,
where \,$\sigma_{\rm SM}^{}=\sigma(pp\to h+{\rm anything})\,{\cal B}(h\to X\bar X)_{\rm SM}$,\,
any change in $\sigma_{\rm SM}^{}$ would also change the limit and hence the disallowed regions.
The presence of the darkon with \,$m_D^{}<m_h^{}/2$\, implies that
${\cal B}(h\to X\bar X)_{\rm SM}$ needs to be replaced by
\,${\cal B}(h\to X\bar X)={\cal R\,B}(h\to X\bar X)_{\rm SM}$\, and
hence $\sigma/\sigma_{\rm SM}^{}$ by~\,$\sigma/(\sigma_{\rm SM}^{}{\cal R})$.\,
This would amount to the weakening of the bounds by a factor~\,$1/\cal R$.

\begin{table}[t] \vspace*{-2ex}
\caption{Ranges of $\cal R$ corresponding to the allowed $m_D^{}$ regions
\,(I)~from 2.5 to 15~GeV\, and \,(II)~not far from, but less than, $m_h^{}/2$,\,
for~\,$m_h^{}=115,150,200,450$~GeV\, in (a)~SM3+D and (b)~SM4+D.}
\begin{tabular}{|c||c|c|c|c|}
\hline
$ $ & \footnotesize 115 & \footnotesize 150 &
\hspace{4.4ex} \footnotesize200 \hspace{4.4ex} & \hspace{4.4ex} \footnotesize450 \hspace{4.4ex} \\
\hline\hline
\scriptsize I\,a  & [0.0007,0.009] & [0.0018,0.020] & [0.058,0.41] & [0.15,0.65] \\
\scriptsize II\,a & [0.56,1] & [0.48,1] & [0.95,1] & [0.97,1] \\
\hline
\scriptsize I\,b  & [0.0014,0.018] & [0.0025,0.029] & [0.065,0.44] & [0.16,0.68] \\
\scriptsize II\,b & [0.79,1] & [0.72,1] & [0.98,1] & [0.99,1] \\
\hline
\end{tabular} \label{rvalues} \vspace*{2ex}
\end{table}

Specifically, the latest 95\%-CL upper-limits on $\sigma/\sigma_{\rm SM}^{}$ in the SM3, and
their median expected limits, reported by ATLAS and CMS~\cite{atlas1,cms1} have minima of
around 0.3 to~0.4.
From Fig.\,\ref{rfactor} and Table~\ref{rvalues}, we can then infer that
the $m_h^{}$ exclusion zone in the SM3, from \,145 to 466~GeV~\cite{atlas1,cms1},
can be entirely recovered in the SM3+D if~\,$m_D^{}\,\mbox{\small$\lesssim$}\,5$\,GeV.\,
Furthermore, the recovered region may become slightly smaller as $m_D^{}$ rises to~15\,GeV.
For \,$m_D^{}\sim m_h^{}/2$,\, with \,${\cal R}\,\mbox{\small$\gtrsim$}\,0.5$,\,
only some of the disallowed Higgs masses can be made viable again.

On a related note, it is intriguing that the modest ({\footnotesize$\sim$}2~standard deviation)
excess at \,$m_h^{}\sim140$\,GeV\, observed by both ATLAS and CMS~\cite{atlas1,cms1} can be
explained as a Higgs having \,${\cal B}_{\rm inv}^{}\sim0.5$\, at
\,$m_D^{}\sim60$\,GeV\, in the SM3+D according to~Fig.\,\ref{rfactor}(a).
A~similar interpretation was previously offered in~Ref.\,\cite{Raidal:2011xk}.

In the SM4 case, the 95\%-CL upper-limits on $\sigma/\sigma_{\rm SM}^{}$
and their median expected limits lie mainly between 0.1 and 0.2, whereas their minima are
roughly \,0.06\,-\,0.08\, and 0.03,\, respectively~\cite{atlas1,cms2}.
Based on Fig.\,\ref{rfactor} and Table~\ref{rvalues}, we then conclude that not all,
but a sizable fraction, of the $m_h^{}$ exclusion zone, 120\,-\,600~GeV, can be recovered
in the SM4+D if \,$m_D^{}\,\mbox{\small$\lesssim$}\,5$\,GeV.\,
As $m_D^{}$ goes~up to 15\,GeV, only \,$m_h^{}\,\mbox{\small$\lesssim$}\,2m_W^{}$\, can be saved.
If~\,$m_D^{}\sim m_h^{}/2$,\, the viable zone is limited to~\,\mbox{$m_h^{}\sim120$\,GeV.}\,
As LHC luminosity grows, the recovered region in the SM4+D may shrink fast,
unless its Higgs is detected.

Finally, the fact that the SM Higgs decay rates into SM particles are not modified by
the darkon's presence implies that the relative sizes of these decay rates in the SM3+D
(SM4+D) are the same as their counterparts in the~SM3~(SM4).
It follows that, if the LHC observes an unambiguous signal of a new electrically neutral
particle in the mass range from \,{\footnotesize$\sim$}115\,GeV to 1\,TeV\, and
if its production rate is below that of the SM Higgs, but still within SM3+D (SM4+D)
expectations, examining the relative rates of the new particle's visible decay modes would be
a~means to help establish whether it is an SM3+D (SM4+D) Higgs or it belongs to some other model.
If it seems to be an SM3+D or SM4+D Higgs, future DM direct searches with improved
sensitivity can check the model further for consistency.

In conclusion, we have explored the implications of the the SM Higgs mass exclusion zones
recently obtained at the LHC for the simplest WIMP DM models, the~SM3+D and~SM4+D.
Current experimental constraints allow $m_D^{}$ values not too far from,
but less than, $m_h^{}/2$ and those compatible with the light-WIMP hypothesis.
In these two regions, the invisible mode \,$h\to DD$\, can dominate the Higgs decay,
with an enhanced branching ratio.
We have demonstrated that, as a consequence, significant portions of the presently excluded
ranges of the SM3 and SM4 Higgs mass may be recovered.
With increased luminosity, the LHC may even uncover a Higgs having SM-like visible decays
still hidden in the currently disallowed regions.
We emphasize that the SM3+D and SM4+D predictions overlap well with the parameter
space for the possible WIMP evidence in the new CRESST-II measurement,
although it is in tension with limits from some other DM experiments.
More precise data from future DM direct searches can test the models more stringently.


\bigskip

This work was supported in part by NSC and NCTS of ROC, NNSF and SJTU 985 grants of PRC, and
NCU Plan to Develop First-Class Universities and Top-Level Research Centers.

\end{document}